\documentclass[article,preprint,groupedaddress]{revtex4}
\usepackage{epsfig}
\usepackage{graphicx}

\begin{document}
\title{Spatially regular charged black holes supporting charged massive scalar clouds}
\author{Shahar Hod}
\affiliation{The Ruppin Academic Center, Emeq Hefer 40250, Israel}
\affiliation{ } \affiliation{The Hadassah Institute, Jerusalem
91010, Israel}
\date{\today}

\begin{abstract}
\ \ \ We prove that, as opposed to the familiar charged Reissner-Nordstr\"om black-hole spacetime, 
the spatially regular charged Ay\'on-Beato-Garc\'ia (ABG) black-hole spacetime can support 
charged scalar clouds, spatially regular stationary matter configurations which are made 
of linearized charged massive scalar fields. 
Interestingly, we reveal the fact that the composed black-hole-field system is 
amenable to an analytical treatment in the regime $Q/M\ll1\ll M\mu$ of weakly charged black holes and large-mass fields, 
in which case it is proved that the dimensionless physical parameter $\alpha\equiv{{qQ}\over{M\mu}}$
must lie in the narrow interval $\alpha\in\big(\sqrt{{{3240}\over{6859}}},{{16}\over{23}}\big)$ 
[here $\{M,Q\}$ are the mass and electric charge of the central black hole 
and $\{\mu,q\}$ are the proper mass and charge coupling constant of the supported scalar field]. 
In particular, we explicitly prove that, for weakly charged black holes, 
the discrete resonance spectrum $\{\alpha(M,Q,\mu,q;n\}^{n=\infty}_{n=0}$ of the 
composed charged-ABG-black-hole-charged-massive-scalar-field cloudy configurations 
can be determined {\it analytically} in the eikonal large-mass regime.   
\end{abstract}
\bigskip
\maketitle


\section{Introduction}

Bosonic field configurations that interact with spinning black holes can be amplified if their 
proper frequencies lie in the superradiant regime
\cite{Zel,PrTe,Viln,Noteunits}
\begin{equation}\label{Eq1}
0<\omega<m\Omega_{\text{H}}\  ,
\end{equation}
where $\Omega_{\text{H}}$ is the horizon angular
velocity of the central black hole and the integer $m$ (with $m>0$) is the azimuthal harmonic index
of the co-rotating bosonic field. 

Interestingly, using analytical techniques in the linearized regime \cite{Hod1,Hod2} 
and numerical computations in the non-linear (self-gravitating) regime \cite{Her1,Her2,Gar1}, 
it has been explicitly proved that the superradiant amplification 
phenomenon may allow spinning black-hole spacetimes to support spatially regular matter configurations which 
are made of stationary minimally-coupled massive bosonic fields which are characterized by 
the critical (marginal) frequency relation \cite{Notemss}
\begin{equation}\label{Eq2}
\omega=m\Omega_{\text{H}}\  .
\end{equation}

It should be emphasized, however, that not all massive bosonic fields can be supported by a central 
spinning black hole which is characterized by a given value $\Omega_{\text{H}}$ of the horizon 
angular velocity. 
In particular, it was proved in \cite{Hodbound} that the mathematically compact black-hole-field 
inequality
\begin{equation}\label{Eq3}
\mu<\sqrt{2}\cdot m\Omega_{\text{H}}\
\end{equation}
provides a necessary condition for the 
existence of composed Kerr-black-hole-massive-scalar-field bound-state configurations, where $\mu$ is the 
proper mass of the supported scalar field. 

As nicely pointed out in \cite{Bekch}, 
a physically analogous phenomenon in which a charged bosonic field is superradiantly amplified by a 
charged black hole occurs if the proper frequency of the incident field lies in the superradiant regime
\begin{equation}\label{Eq4}
0<\omega<q\Phi_{\text{H}}\  ,
\end{equation}
where $\Phi_{\text{H}}$ is the electric potential at the outer horizon of the central 
charged black hole and $q$ is the charge coupling constant of the field \cite{Bekch}.

Intriguingly, however, it has been explicitly proved in \cite{Hodnbt} that the canonical charged 
Reissner-Nordstr\"om (RN) black-hole spacetime cannot support spatially regular linearized matter 
configurations which are made of minimally coupled (static or stationary) charged massive scalar fields. 
In particular, it has been revealed in \cite{Hodnbt} that, 
as opposed to the case of spinning black holes \cite{Hod1,Hod2,Her1,Her2,Gar1}, 
the mutual gravitational attraction between a central charged RN black hole and a 
charged massive scalar field cannot provide, in the superradiant regime (\ref{Eq4}), the confinement mechanism 
which is required in order to prevent the scalar field from radiating its energy to infinity \cite{Notecbb,Dego,Herrec,Hodexp,Lim}. 

It has recently been demonstrated numerically in the physically interesting paper \cite{PL} that charged scalar fields 
can be superradiantly amplified in the charged Ay\'on-Beato-Garc\'ia (ABG) \cite{ABG} spacetime 
which describes a spatially regular 
black-hole solution of the coupled Einstein-non-linear-electrodynamics field equations. 
In particular, it has been intriguingly suggested in \cite{PL} that 
the charged ABG black-hole spacetime may be superradiantly unstable to linearized 
perturbations of charged massive scalar fields. 

The main goal of the present paper is to explore, using {\it analytical} techniques, the 
physical and mathematical properties of the 
composed charged-ABG-black-hole-charged-massive-scalar-field system. 
Interestingly, we shall explicitly prove below that, as opposed to the charged RN black-hole spacetime, 
the spatially regular charged ABG black-hole spacetime can support stationary scalar clouds, 
spatially regular matter configurations which are made of linearized charged massive scalar fields. 
These composed black-hole-linearized-scalar-field stationary bound-state configurations, which are 
characterized by the critical (marginal) frequency relation $\omega=q\Phi_{\text{H}}$ 
for the superradiant amplification phenomenon in the charged black-hole spacetime, 
mark the onset of the superradiant instabilities in the charged ABG black-hole spacetime to perturbations of 
charged massive scalar fields in the frequency regime (\ref{Eq4}).

Interestingly, in the present compact paper we shall explicitly prove that the composed 
charged-ABG-black-hole-charged-massive-scalar-field system is 
amenable to an analytical treatment in the dimensionless regime $Q/M\ll1\ll M\mu$ of weakly charged black holes 
and large-mass fields \cite{NoteMQmu}.

\section{Description of the system}

We consider a physical system which is composed of a charged massive scalar field of proper mass $\mu$ and 
electric charge $q$ which is 
linearly coupled to a central charged Ay\'on-Beato-Garc\'ia black hole \cite{ABG} of mass $M$ 
and electric charge $Q$ \cite{Noteqp,Notequh}. 
The ABG spacetime describes a spatially regular 
solution of the coupled Einstein-non-linear-electrodynamics field equations (see \cite{ABG,PL} and references 
therein for details). 

The black-hole spacetime is described, using the familiar Schwarzschild 
coordinates $\{t,r,\theta,\varphi\}$, by the curved line element \cite{Chan}
\begin{equation}\label{Eq5}
ds^2=-f(r)dt^2+{1\over{f(r)}}dr^2+r^2(d\theta^2+\sin^2\theta d\varphi^2)\ ,
\end{equation}
where the metric function in (\ref{Eq5}) is given by the radially-dependent expression \cite{ABG,PL}
\begin{equation}\label{Eq6}
f^{\text{ABG}}(r)=1-{{2Mr^2}\over{(r^2+Q^2)^{3/2}}}+{{Q^2r^2}\over{(r^2+Q^2)^2}}\  .
\end{equation}
The horizon radii of the charged ABG spacetime are determined by the zeros of the 
metric function $f(r)$:
\begin{equation}\label{Eq7}
f(r)=0\ \ \ \ \text{for}\ \ \ \ r=r_{\text{H}}\  . 
\end{equation}
The radially-dependent electrostatic potential of the ABG spacetime (\ref{Eq5}) is given by the non-trivial 
functional expression \cite{ABG,PL}
\begin{equation}\label{Eq8}
\Phi^{\text{ABG}}(r)={{r^5}\over{2Q}}\Big[{{3M}\over{r^5}}+{{2Q^2}\over{(r^2+Q^2)^3}}-{{3M}
\over{(r^2+Q^2)^{5/2}}}\Big]\  .
\end{equation}

The dynamics of a linearized charged massive scalar field $\Psi$ in the curved black-hole spacetime (\ref{Eq5}) 
is governed by the familiar Klein-Gordon wave equation \cite{PL,HodPirpam,Stro,HodCQG2}
\begin{equation}\label{Eq9}
[(\nabla^\nu-iqA^\nu)(\nabla_{\nu}-iqA_{\nu}) -\mu^2]\Psi=0\  ,
\end{equation}
where $A_{\nu}=-\delta_{\nu}^{0}\Phi(r)$ is the electromagnetic
potential [see Eq. (\ref{Eq8})] of the charged black hole. 
Using the field decomposition 
\begin{equation}\label{Eq10}
\Psi(t,r,\theta,\varphi)=\int\sum_{lm} {{1}\over{r}}R_{lm}(r)Y_{lm}(\theta,\varphi)e^{im\varphi}e^{-i\omega t}d\omega\
\end{equation}
for the scalar wave function, 
where $Y_{lm}(\theta,\varphi)$ are the familiar spherical harmonic functions (with $l\geq|m|$ \cite{Notelm}), 
and using the tortoise radial coordinate $y(r)$, which is defined by the differential relation \cite{Notery} 
\begin{equation}\label{Eq11}
dy={{dr}\over{f(r)}}\  ,
\end{equation}
one finds from Eqs. (\ref{Eq5}), (\ref{Eq6}), and (\ref{Eq9}) the Schr\"odinger-like ordinary differential equation \cite{Noteom}
\begin{equation}\label{Eq12}
{{d^2R}\over{dy^2}}-VR=0\
\end{equation}
for the scalar field. 
The effective potential $V[r(y)]$ of the composed 
charged-ABG-black-hole-charged-massive-scalar-field system is given by the radially-dependent 
functional expression \cite{PL}
\begin{eqnarray}\label{Eq13}
V[r(y)]=f(r)\Big[\mu^2+{{1}\over{r}}{{df(r)}\over{dr}}+{{l(l+1)}\over{r^2}}\Big]-
\big[\omega-q\Phi(r)\big]^2\  .
\end{eqnarray}

The differential equation (\ref{Eq12}) of the charged massive scalar field is supplemented by the 
large-$r$ boundary condition \cite{Hodnbt}
\begin{equation}\label{Eq14}
R \sim e^{-\sqrt{\mu^2-\omega^2}y}\ \ \ \ \text{for}\ \ \ \ r\to\infty\ \ (y\rightarrow \infty)\
\end{equation}
which, in the bounded frequency regime 
\begin{equation}\label{Eq15}
\omega^2<\mu^2\  ,
\end{equation}
characterizes normalizable (spatially bounded) scalar eigenfunctions. 
In addition, the inner boundary condition \cite{Noterh} 
\begin{equation}\label{Eq16}
R \sim e^{-i (\omega-q\Phi_{\text{H}})y}\ \ \ \ \text{for}\ \ \ \ r\to r_{\text{H}}\ \ (y\rightarrow -\infty)\
\end{equation}
for the scalar field, where [see Eqs. (\ref{Eq8}) and (\ref{Eq10})]
\begin{equation}\label{Eq17}
\Phi_{\text{H}}\equiv\Phi(r=r_{\text{H}})\  ,
\end{equation}
describes purely ingoing waves at the outer horizon of the central black hole (as measured by a comoving observer). 

Interestingly, in the next section we shall explicitly prove that 
the set of equations (\ref{Eq12}), (\ref{Eq13}), (\ref{Eq14}), and (\ref{Eq16}) 
determine the discrete resonance spectrum of the composed 
charged-ABG-black-hole-charged-massive-scalar-field cloudy configurations. 

\section{The discrete resonance spectrum of the composed ABG-black-hole-linearized-charged-massive-scalar-field cloudy 
configurations}

In the present section we shall reveal the fact that the physical and mathematical properties of the 
composed charged-ABG-black-hole-linearized-charged-massive-scalar-field system can be studied {\it analytically} 
in the dimensionless regime 
\begin{equation}\label{Eq18}
{{Q}\over{M}}\ll1\ll M\mu\ll Mq\  .
\end{equation}
The strong inequalities (\ref{Eq18}) characterize weakly charged ABG black holes and massive scalar fields whose
Compton wavelengths are much smaller than the characteristic lengthscale $M$ which is set by the radius 
of the central black hole. 

In particular, we shall explicitly prove that the discrete resonance spectrum of the 
dimensionless charge-mass parameter 
\begin{equation}\label{Eq19}
\alpha\equiv {{qQ}\over{M\mu}}\  ,
\end{equation}
which characterizes the composed black-hole-scalar-field bound-state configurations, 
can be determined analytically in the regime (\ref{Eq18}).  

The stationary (marginally-stable) bound-state scalar configurations in the ABG black-hole spacetime (\ref{Eq5}) 
are characterized by the critical frequency 
\begin{equation}\label{Eq20}
\omega=q\Phi_{\text{H}}\
\end{equation}
for the superradiant amplification phenomenon of bosonic fields in the charged black-hole spacetime. 
In particular, taking cognizance of Eq. (\ref{Eq16}) one deduces that, for scalar fields 
with the critical frequency (\ref{Eq20}), 
there is no net flux of energy through the outer horizon of the central supporting black hole.

Taking cognizance of the strong inequalities (\ref{Eq18}), 
the metric function (\ref{Eq6}) can be approximated by
\begin{equation}\label{Eq21}
f(r)=1-{{2M}\over{r}}+O(Q^2/r^2)\  ,
\end{equation}
which yields the expression [see Eq. (\ref{Eq7})]
\begin{equation}\label{Eq22}
r_{\text{H}}=2M\cdot[1+O(Q^2/M^2)]\
\end{equation}
for the radius of the black-hole outer horizon.

In addition, from Eqs. (\ref{Eq8}) and (\ref{Eq18}) one finds the relation
\begin{equation}\label{Eq23}
\Phi(r)={{Q}\over{r}}\cdot\Big[1+{{15M}\over{4r}}+O(Q^2/r^2)\Big]\
\end{equation}
for the black-hole electric potential, which yields the simple horizon relation [see Eq. (\ref{Eq22})]
\begin{equation}\label{Eq24}
\Phi_{\text{H}}={{23}\over{16}}\cdot{{Q}\over{M}}\cdot[1+O(Q^2/M^2)]\  .
\end{equation}

Substituting Eqs. (\ref{Eq19}), (\ref{Eq20}), (\ref{Eq21}), (\ref{Eq23}), and (\ref{Eq24}) into Eq. (\ref{Eq13}), 
one finds that the composed black-hole-field radial potential can be approximated by 
\begin{eqnarray}\label{Eq25}
V(r)=\mu^2\cdot\Big(1-{{2M}\over{r}}\Big)\Big[1-\alpha^2\cdot
\Big(1-{{2M}\over{r}}\Big)\Big({{23r+30M}\over{16r}}\Big)^2\Big]\  .
\end{eqnarray}

\subsection{Upper and lower bounds on the allowed values of the dimensionless physical parameter $\alpha$}

In the present subsection we shall derive two necessary conditions for the existence of composed 
charged-ABG-black-hole-charge-field bound-state configurations in the dimensionless regime (\ref{Eq18}). 

To this end, we first point out that Eqs. (\ref{Eq15}), (\ref{Eq20}), and (\ref{Eq24}) imply that, 
in the regime (\ref{Eq18}), spatially bounded scalar configurations are characterized by the dimensionless inequality
\begin{equation}\label{Eq26}
\Big({{23}\over{16}}\cdot{{qQ}\over{M}}\Big)^2<\mu^2\  ,
\end{equation}
which yields the upper bound [see Eq. (\ref{Eq19})]
\begin{equation}\label{Eq27}
\alpha<{{16}\over{23}}\  .
\end{equation}

In addition, we point out that the radial potential (\ref{Eq25}) is characterized by the asymptotic 
functional behaviors [see Eq. (\ref{Eq22})] \cite{Noteann}
\begin{equation}\label{Eq28}
V(r\to r^+_{\text{H}})\to0^+\
\end{equation}
and 
\begin{equation}\label{Eq29}
V(r\to\infty)\to \mu^2\cdot\Big[1-\Big({{23\alpha}\over{16}}\Big)^2\Big]>0\  .
\end{equation}
The existence of a binding potential well outside the black-hole horizon provides a 
necessary condition for the existence of stationary bound-state configurations of the 
charged massive scalar fields in the curved spacetime (\ref{Eq5}). 
In particular, taking cognizance of the asymptotic properties (\ref{Eq28}) and (\ref{Eq29}) of the composed black-hole-field 
radial potential, one deduces that the requirement
\begin{equation}\label{Eq30}
\text{min}_r\{V(r)\}<0\
\end{equation}
provides a necessary condition for the existence of composed ABG-black-hole-scalar-field 
bound-state configurations. 

The dimensionless function [see Eq. (\ref{Eq25})]
\begin{equation}\label{Eq31}
{\cal F}(r)\equiv1-\alpha^2\cdot\Big(1-{{2M}\over{r}}\Big)\Big({{23r+30M}\over{16r}}\Big)^2\
\end{equation}
is characterized by the radial minimum (for $r\geq r_{\text{H}}$)
\begin{equation}\label{Eq32}
\text{min}_r\{{\cal F}(r)\}=1-{{6859}\over{3240}}\cdot\alpha^2\ \ \ \ \text{for}\ \ \ \ r_{\text{min}}={{90}\over{7}}M\  .
\end{equation}
Thus, from Eqs. (\ref{Eq25}), (\ref{Eq30}), and (\ref{Eq32}) one finds that the dimensionless 
lower bound 
\begin{equation}\label{Eq33}
\alpha>\sqrt{{{3240}\over{6859}}}\
\end{equation}
provides a necessary condition for the existence of 
composed ABG-black-hole-scalar-field bound-state configurations.

Taking cognizance of the analytically derived necessary conditions (\ref{Eq27}) and (\ref{Eq33}) one deduces that, 
in the regime (\ref{Eq18}), the 
dimensionless physical parameter $\alpha$ which characterizes the 
composed charged-ABG-black-hole-charged-massive-scalar-field cloudy configurations 
must lie in the narrow \cite{Noteul} interval 
\begin{equation}\label{Eq34}
\sqrt{{{3240}\over{6859}}}<\alpha<{{16}\over{23}}\  .
\end{equation}

\subsection{The resonance spectrum of the composed charged-ABG-black-hole-charged-massive-scalar-field cloudy configurations}

In the present subsection we shall analyze the resonance spectrum 
$\{\alpha(M,Q,\mu,q;n\}^{n=\infty}_{n=0}$ of the dimensionless charge-mass parameter which 
characterizes the composed ABG-black-hole-charged-massive-scalar-field system. 
Interestingly, we shall explicitly prove that the discrete resonance spectrum can be determined {\it analytically} 
in the near-critical regime [see Eq. (\ref{Eq33})]
\begin{equation}\label{Eq35}
\alpha\gtrsim\sqrt{{{3240}\over{6859}}}\  .
\end{equation}

In particular, the Schr\"odinger-like radial differential equation (\ref{Eq12}) of the supported charged massive scalar fields 
in the charged ABG black-hole spacetime (\ref{Eq5}) is characterized by the well known 
second-order WKB quantization condition \cite{WKB1,WKB2,WKB3}
\begin{equation}\label{Eq36}
\int_{y_{t_-}}^{y_{t_+}}dy\sqrt{-V(y;M,Q,\mu,q)}=\big(n+{1\over2}\big)\cdot\pi\
\ \ \ ; \ \ \ \ n=0,1,2,...\  ,
\end{equation}
where the integration limits $\{y_{t_-},y_{t_+}\}$, which are determined by the radial relations
\begin{equation}\label{Eq37}
V(y_{t_-})=V(y_{t_+})=0\  ,
\end{equation}
are the classical turning points of the binding potential (\ref{Eq25}). 
The integer $n\in\{0,1,2,...\}$ in the WKB integral relation (\ref{Eq36}) is the discrete 
resonance parameter of the composed black-hole-field system. 
Taking cognizance of the differential relation (\ref{Eq11}), one can express the WKB resonance 
condition (\ref{Eq36}) in the form
\begin{equation}\label{Eq38}
\int_{r_{t_-}}^{r_{t_+}}dr \sqrt{-{{V(r;M,Q,\mu,q)}\over{[f(r)]^2}}}=\big(n+{1\over2}\big)\cdot\pi\
\ \ \ ; \ \ \ \ n=0,1,2,...\  .
\end{equation}

Using the dimensionless physical variables $\{\epsilon,x\}$, which are defined by the relations 
[see Eqs. (\ref{Eq32}) and (\ref{Eq35})]
\begin{equation}\label{Eq39}
\alpha\equiv\sqrt{{{3240}\over{6859}}}\cdot(1+\epsilon)\ \ \ \ ; \ \ \ \ 0\leq\epsilon\ll1\
\end{equation}
and
\begin{equation}\label{Eq40}
r\equiv r_{\text{min}}\cdot(1+x)\ \ \ \ ; \ \ \ \ |x|\ll1\  ,
\end{equation}
one can write the effective radial potential (\ref{Eq25}) of the composed 
charged-ABG-black-hole-charged-massive-scalar-field configurations 
in the dimensionless form \cite{Notehu}
\begin{equation}\label{Eq41}
V(x)=\mu^2\cdot{{38}\over{45}}\cdot\Big(-2\epsilon+{{147}\over{5776}}\cdot x^2\Big)
\cdot[1+O(x,\epsilon)]\  .
\end{equation}

Substituting Eqs. (\ref{Eq40}) and (\ref{Eq41}) into Eq. (\ref{Eq38}) and defining the 
dimensionless variable
\begin{equation}\label{Eq42}
z=\sqrt{{{147}\over{11552\epsilon}}}\cdot x\  ,
\end{equation}
one obtains the remarkably compact WKB resonance condition
\begin{equation}\label{Eq43}
\epsilon\cdot M\mu\cdot{{360\sqrt{570}}\over{49}}
\int_{-1}^{1}dz \sqrt{1-z^2}=\big(n+{1\over2})\cdot\pi\ \ \ \ ; \ \ \
\ n=0,1,2,...\    
\end{equation}
for the composed charged-black-hole-charged-massive-scalar-field cloudy configurations. 
Performing the integration in (\ref{Eq43}), one finds the discrete resonance spectrum \cite{Noteintegral}
\begin{equation}\label{Eq44}
\epsilon_n={{49}\over{180\sqrt{570}M\mu}}\cdot\big(n+{1\over2})\ \ \ \ ; \ \ \ \ n=0,1,2,...\  .
\end{equation}
As a consistency check we point out that one deduces from Eq. (\ref{Eq44}) the strong 
inequality $\epsilon\ll1$ in the large-mass $M\mu\gg1$ 
regime [see Eqs. (\ref{Eq18}) and (\ref{Eq39})]. 

Substituting the analytically derived relation (\ref{Eq44}) into Eq. (\ref{Eq39}), one obtains the 
discrete large-mass resonance spectrum
\begin{equation}\label{Eq45}
\alpha_n=\sqrt{{{3240}\over{6859}}}+{{49\sqrt{3}}\over{10830M\mu}}\cdot\big(n+{1\over2})
\ \ \ \ ; \ \ \ \ n=0,1,2,...\
\end{equation}
of the composed charged-ABG-black-hole-charged-massive-scalar-field cloudy configurations.

\section{The effective radial widths of the supported charged massive scalar clouds}

In the present section we shall determine the effective radial widths of the 
stationary charged massive scalar clouds that are supported in the charged ABG 
black-hole spacetime (\ref{Eq5}). 
In particular, we shall reveal the physically interesting fact that, in the dimensionless 
large-mass $M\mu\gg1$ regime [see Eq. (\ref{Eq18})], the supported scalar configurations 
can be made arbitrarily thin. 

The effective widths of the supported charged scalar clouds in the charged ABG black-hole spacetime 
are determined by the classically allowed radial region [see Eq. (\ref{Eq37})]
\begin{equation}\label{Eq46}
\Delta r(M,Q,\mu,q)\equiv r_{t_+}-r_{t_-}\
\end{equation}
of the composed black-hole-field binding potential (\ref{Eq25}). 
In particular, from Eq. (\ref{Eq41}) one finds the simple functional relation
\begin{equation}\label{Eq47}
\Delta x(M,Q,\mu,q)\equiv x_{t_+}-x_{t_-}=\sqrt{{{46208}\over{147}}}\cdot\sqrt{\epsilon}\  ,
\end{equation}
which, taking cognizance of Eqs. (\ref{Eq32}) and (\ref{Eq40}), yields the expression
\begin{equation}\label{Eq48}
{{\Delta r}\over{M}}={{4560\sqrt{6}}\over{49}}\cdot\sqrt{\epsilon}\
\end{equation}
for the effective dimensionless widths of 
the charged massive scalar field configurations that are supported in the charged ABG 
black-hole spacetime (\ref{Eq5}). 

Substituting the resonance relation (\ref{Eq44}) into Eq. (\ref{Eq48}), one obtains the dimensionless expression
\begin{equation}\label{Eq49}
{{\Delta r}\over{M}}={{8\sqrt{19}\sqrt[4]{570}}\over{7}}
\cdot\sqrt{n+{1\over2}}\cdot{{1}\over{\sqrt{M\mu}}}\
\end{equation}
for the effective widths of the scalar clouds. 
From the analytically derived functional expression (\ref{Eq49}) one finds that the stationary charged 
scalar clouds, which are supported in the charged ABG black-hole spacetime (\ref{Eq5}), 
can be made arbitrarily thin in the dimensionless large-mass $M\mu\gg1$ regime \cite{Notecq}. 

\section{Summary and discussion}

It has recently been demonstrated in the physically important paper \cite{PL} that charged 
scalar fields can be superradiantly amplified in the charged ABG black-hole spacetime \cite{ABG} 
which describes a spatially regular 
solution of the coupled Einstein-non-linear-electrodynamics field equations.

Motivated by the interesting numerical results presented in \cite{PL}, we have studied, using analytical techniques, 
the physical and mathematical properties of charged massive scalar field configurations 
(stationary scalar clouds) that are supported by charged ABG black holes. 
In particular, we have explicitly proved that the composed charged-ABG-black-hole-charged-massive-scalar-field 
system can be studied {\it analytically} in the dimensionless regime $Q/M\ll1\ll M\mu$ of weakly charged black holes 
and large-mass fields. 

The main analytical results derived in this paper and their physical implications are as follows:

(1) We have proved that, in the regime (\ref{Eq18}), the dimensionless 
physical parameter $\alpha\equiv{{qQ}/{M\mu}}$, which characterizes the composed 
charged-black-hole-charged-field bound-state 
configurations, must lie in the narrow interval [see Eqs. (\ref{Eq19}) and (\ref{Eq34})]
\begin{equation}\label{Eq50}
{{qQ}\over{M\mu}}\in\Big(\sqrt{{{3240}\over{6859}}},{{16}\over{23}}\Big)\  .
\end{equation}

(2) Using a WKB analysis in the eikonal large-mass regime (\ref{Eq18}), 
we have derived the remarkably compact analytical resonance formula [see Eqs. (\ref{Eq19}) and (\ref{Eq45})]
\begin{equation}\label{Eq51}
\Big({{qQ}\over{M\mu}}\Big)_n=
\sqrt{{{3240}\over{6859}}}+{{49\sqrt{3}}\over{10830M\mu}}\cdot\big(n+{1\over2})
\ \ \ \ ; \ \ \ \ n=0,1,2,...\
\end{equation}
for the dimensionless charge-mass parameter $qQ/M\mu$ which characterizes 
the composed charged-ABG-black-hole-charged-massive-scalar-field bound-state configurations. 

(3) We have proved that the stationary charged massive scalar clouds 
are characterized by the effective dimensionless widths \cite{Noten01}
\begin{equation}\label{Eq52}
{{\Delta r}\over{M}}={{4\sqrt{38}\sqrt[4]{570}}\over{7}}
\cdot{{1}\over{\sqrt{M\mu}}}\  .
\end{equation}
Interestingly, the analytically derived functional expression (\ref{Eq52}) reveals the fact 
that the charged scalar configurations, which are supported in the charged ABG black-hole 
spacetime (\ref{Eq5}), are extremely thin (with $\Delta r/M\ll1$) 
in the dimensionless large-mass $M\mu\gg1$ regime.  

(4) It is worth emphasizing that the analytically derived critical existence-line \cite{Noten02}
\begin{equation}\label{Eq53}
{{qQ}\over{M\mu}}=
\sqrt{{{3240}\over{6859}}}+{{49\sqrt{3}}\over{21660M\mu}}\  ,
\end{equation}
which characterizes the stationary charged-ABG-black-hole-charged-massive-scalar-field bound-state configurations, marks 
in the dimensionless large-mass regime (\ref{Eq18}) the sharp 
boundary between bald ABG black-hole spacetimes 
and charged black holes that are superradiantly unstable to perturbations of 
charged massive scalar fields. 

In particular, the expected onset of superradiant instabilities in the composed ABG-black-hole-scalar-field 
system above the critical existence-line (\ref{Eq53}) hints that this physically 
interesting system may be characterized by the existence of charged hairy black-hole configurations 
that support non-linear (self-gravitating) charged massive scalar fields. 

Finally, we would like to stress again that, in the present analysis, the minimally coupled charged massive scalar 
fields were treated at the linearized level. 
As we explicitly shown, the main advantage of this approach lies in the fact that the composed 
charged-ABG-black-hole-charged-massive-linearized-scalar-field system is amenable to an {\it analytical} treatment. 
We believe that it would be highly interesting (and physically important) to use non-linear {\it numerical} techniques 
in order to prove the existence of genuine spatially regular hairy (scalarized) charged ABG black-hole spacetimes. 

\bigskip
\noindent
{\bf ACKNOWLEDGMENTS}
\bigskip

This research is supported by the Carmel Science Foundation. I thank
Yael Oren, Arbel M. Ongo, Ayelet B. Lata, and Alona B. Tea for
stimulating discussions.



\begin{thebibliography}{99}
\bibitem{Noteunits} We use natural units in which $G=c=\hbar=1$.
\bibitem{Notemss} Note that stationary field modes, which are characterized by the critical 
frequency relation (\ref{Eq2}) for the superradiant scattering
phenomenon in the spinning black-hole spacetime, mark the boundary between stable ($\Im\omega<0$) and
unstable ($\Im\omega>0$) resonances of the massive bosonic fields \cite{Hod1,Hod2,Her1,Her2,Gar1}.
\bibitem{Notecbb} It is physically interesting to note that the charged 
RN black hole may become superradiantlly unstable to charged perturbations 
of bosonic fields if it is surronded by a reflecting surface (a mirror) that prevents the 
charged bosonic field from radiating its energy to infinity \cite{Dego,Herrec,Hodexp,Lim}.
\bibitem{NoteMQmu} Here $M$ and $Q$ are respectively the mass and electric charge of the ABG 
black hole and $\mu$ is the proper mass of the scalar field.
\bibitem{Noteqp} We shall assume, without loss of generality, the relations $q>0$ and $Q>0$ for the 
electric charges of the scalar field and the central black hole.
\bibitem{Notequh} Note that the physical parameters $q$ and $\mu$ of the charged massive 
scalar field stand respectively for $q/\hbar$ and $\mu/\hbar$ and therefore 
have the dimensions of $($length$)^{-1}$.
\bibitem{Notelm} Here the integers $l$ and $m$ 
are respectively the spherical harmonic index and the azimuthal harmonic index which characterize 
the scalar field mode.
\bibitem{Notery} Note that the differential relation (\ref{Eq11}) maps the radial coordinate 
$r\in[r_{\text{H}},\infty]$ to the radial coordinate $y\in[-\infty,\infty]$.
\bibitem{Noteom} For brevity, we shall henceforth omit the angular harmonic indices $\{l,m\}$ of the 
charged massive scalar field.
\bibitem{Noterh} We shall henceforth use the notation $r_{\text{H}}$ for the radius of the 
black-hole outer horizon.
\bibitem{Noteann} It is worth noting that the exact radial potential (\ref{Eq13}) is also characterized 
by the horizon behavior (\ref{Eq28}) [see Eqs. (\ref{Eq7}), (\ref{Eq17}), and (\ref{Eq20})].
\bibitem{Noteul} Interestingly, one finds that the upper bound $16/23$ on the value of the 
dimensionless physical parameter $\alpha$ 
is larger than the lower bound $\sqrt{{{3240}/{6859}}}$ by only $\sim1\%$ [see Eq. (\ref{Eq34})].
\bibitem{Notehu} Here we have used the relation $f(r)=({{38}/{45}})\cdot[1+O(x)]$ 
[see Eqs. (\ref{Eq21}) and (\ref{Eq32})].
\bibitem{Noteintegral} Here we have used the integral relation 
$\int_{-1}^{1}dz \sqrt{1-z^2}={{\pi}/{2}}$.
\bibitem{Notecq} It is worth noting that the charged massive scalar field is exponentially 
suppressed outside the classically allowed narrow radial interval (\ref{Eq49}).
\bibitem{Noten01} Here we have used the dimensionless functional relation (\ref{Eq49}) with the value 
$n=0$ for the fundamental resonant mode of the 
composed charged-black-hole-charged-scalar-field system.
\bibitem{Noten02} Here we have used the analytically derived resonance spectrum (\ref{Eq51}) with the value 
$n=0$ for the fundamental resonant mode of the composed charged-black-hole-charged-scalar-field system.
\end{thebibliography}
\end{document}